# Can two chaotic systems give rise to order?

J. Almeida[1], D. Peralta-Salas [2*], and M. Romera[3]

[1]*Departamento de Física Teórica I, Facultad de Ciencias Físicas, Universidad Complutense, 28040 Madrid (Spain).*

[2]*Departamento de Física Teórica II, Facultad de Ciencias Físicas, Universidad Complutense, 28040 Madrid (Spain).*

[3]*Instituto de Física Aplicada, Consejo Superior de Investigaciones Científicas, Serrano 144, 28006 Madrid (Spain).*

The recently discovered Parrondo's paradox claims that two losing games can result, under random or periodic alternation of their dynamics, in a winning game: "losing＋losing＝winning". In this paper we follow Parrondo's philosophy of combining different dynamics and we apply it to the case of one-dimensional quadratic maps. We prove that the periodic mixing of two chaotic dynamics originates an ordered dynamics in certain cases. This provides an explicit example (theoretically and numerically tested) of a different Parrondian paradoxical phenomenon: "chaos＋chaos ＝ order".

## 1. Introduction

One of the most extended ways to model the evolution of a natural process (physical, biological or even economical) is by employing discrete dynamics, that is, maps which apply one point to another point of certain variables space. Deterministic or random laws are allowed, and in this last case the term game is commonly used instead of map (physicists also use the term discrete random process). Let us assume that we have two different discrete dynamics $A_1$ and $A_2$. In the last decades a great effort has been done in understanding at least the most significant qualitative aspects of each dynamics separately. A different (but related) topic of research which has arisen in the last years consists in studying the dynamics obtained by combination of the dynamics $A_1$ and $A_2$:

$$x_0 \xrightarrow{A_{H_0}} x_1 \xrightarrow{A_{H_1}} x_2 \xrightarrow{A_{H_2}} x_3 \ \ldots$$

where $H$ stands for certain deterministic or random law which assign the value 1 or 2 to each number of the sequence $\{0,1,2,...\}$ and $\{x_0, x_1, x_2,...\}$ are the values of the variable $x$ describing the physical system.

To the best of our knowledge the idea of alternating different dynamics and comparing certain properties of the combined dynamics with the properties of the individual dynamics is due to Parrondo and collaborators [1-6]. They constructed two simple games with negative gains (they were losing games) but when they were alternated in different (random or deterministic) manners the gain was positive (the resulting game was winning). By the term gain we mean the asymptotic gain per move as defined, for instance, in reference [7]. This apparent contradiction is known as Parrondo's paradox and since its discovery it has become an active area of research.

Let us briefly review the original games $A_1$ and $A_2$ discovered by Parrondo

Game $A_1$: $\begin{cases} \boldsymbol{r}(x \rightarrow x+1) = p \\ \boldsymbol{r}(x \rightarrow x-1) = 1-p \ , \end{cases}$

Game $A_2$: $\begin{cases} \begin{cases} \boldsymbol{r}(x \rightarrow x+1) = p_1 \\ \boldsymbol{r}(x \rightarrow x-1) = 1-p_1 \end{cases} & \text{if } |x| \bmod 3 = 0 \\ \begin{cases} \boldsymbol{r}(x \rightarrow x+1) = p_2 \\ \boldsymbol{r}(x \rightarrow x-1) = 1-p_2 \end{cases} & \text{if } |x| \bmod 3 \neq 0 \ , \end{cases}$

where $\boldsymbol{r}(i \rightarrow j)$ stands for the transition probability and the variable $x$ can only take integer values. It can be shown (see references [1-6] or reference [7] for a different approach) that games $A_1$ and $A_2$ are losing whenever the following conditions hold:

$$p < \frac{1}{2},$$
$$f \equiv \frac{(1-p_1)(1-p_2)^2}{p_1 p_2^2} > 1 \ . \tag{1}$$

In the same references it is proved that the game obtained by random combination of games $A_1$ and $A_2$ (with probability $\sigma$ of playing with game $A_1$) is winning if

$$\Lambda \equiv \frac{\left(1-\boldsymbol{s}\,p-(1-\boldsymbol{s})p_1\right)\left(1-\boldsymbol{s}\,p-(1-\boldsymbol{s})p_2\right)^2}{\left(\boldsymbol{s}\,p+(1-\boldsymbol{s})p_1\right)\left(\boldsymbol{s}\,p+(1-\boldsymbol{s})p_2\right)^2} < 1 \ . \tag{2}$$

Parrondo's paradox occurs for values of the parameters $(p, p_1, p_2, \boldsymbol{s})$ satisfying the inequalities of Eqs. (1) and (2), for instance $p = 5/11$, $p_1 = 1/21$, $p_2 = 10/11$ and $\boldsymbol{s} = 1/2$. Furthermore, since Eqs. (1) and (2) define an open set in the parameters space, there are infinite values for which the paradox occurs. See Fig. 1 where we have represented in the cube $(0 \leq p_1 \leq 1 , 0 \leq p_2 \leq 1 , 0 \leq p \leq 0.5)$ the region for which inequalities of Eqs. (1) and (2) hold ($\sigma$ is fixed to the value 0.4).

Deterministic combinations of games $A_1$ and $A_2$ also yield paradoxical phenomena. In Fig. 2 we have represented the expected gain per move for different alternations of games. The numerical simulations were averaged over 50,000 trials. The interested reader can see references [1-7] for a more detailed study of the original Parrondo's paradox.

In this paper however we are not interested in the paradoxical phenomenon "losing + losing = winning" but in extending Parrondo's philosophy to the combination of two non linear deterministic dynamics. We will study the Parrondian phenomenon arising when two dynamics $A_1$ and $A_2$ experiment a dramatic change of their properties when they are combined, as happens in the original Parrondo's paradox for games.

Specifically we will consider the one-dimensional quadratic map

$$x_{n+1} = x_n^2 + c ,$$ (3)

and we will show in the following section that for certain values of the parameter $c$ the dynamics $A_1 : x_{n+1} = x_n^2 + c_1$ and $A_2 : x_{n+1} = x_n^2 + c_2$ are chaotic but the dynamics obtained by periodic alternation $A_1 A_2 A_1 A_2 \ldots \equiv (A_1 A_2)$ is ordered in a well defined sense. This is a new Parrondian phenomenon which can be stated as "chaos + chaos = order" by analogy with the original Parrondo's paradox. We have not been able to find examples of "chaos + chaos = order" in the literature, at least as clear as the examples we provide in this paper. We are not aware that this aspect of the theory of "iterated function systems" has previously been investigated. In fact the only references that we have found relating Parrondo's paradox and chaos are the works of Bucolo [8] and Arena [9]. Anyway our results have nothing to do with these papers since they consider the standard Parrondo's paradox "losing + losing = winning" and introduce chaotic maps in the way of combining the games but they never study the phenomenon "chaos + chaos = order".

Since chaos is closely linked to instability and fractal structures it is worth mentioning the interesting works of Allison and collaborators [10], [11]. In the first one it is designed a switched mode circuit which is unstable in either mode $A_1, A_2$ but is stable when switched at random. The single evolutions are linear and the dynamics $A_1$ and $A_2$ possess a saddle point and an unstable focus respectively. In contrast the switched dynamics has a stable node. In the second paper it is studied the time evolution of the probability vectors according to a chain of Markov operators. When this evolution is homogeneous then the state-space representation converges to a single point but when the evolution is given by a random sequence of two Markov operators the attractor in the state-space is a fractal set (in fact a Cantor type set). A chaotic system has two main ingredients, instability and ergodicity in compact domains, so the phenomenon that we report is stronger than the idea "instability+ instability=stability" studied in reference [10], in which the single dynamics are unstable but not chaotic because they are linear. On the other hand in reference [11] the single dynamics are also linear, in strong contrast with our quadratic systems. Furthermore the existence of a fractal attractor is not necessarily related to chaos, there exist strange non-chaotic attractors [12]. The Lyapounov exponents must be positive, a fact which is not checked in reference [11] for the randomly switched dynamics.

Note that "chaos" and "losing" are unrelated concepts. The link between the original Parrondo's paradox and the phenomenon "chaos+ chaos=order" reported in this paper is that both are phenomena in which one property of the single dynamics ("losing" and "chaos" respectively) completely changes when alternating them. We call these Parrondian phenomena. Note the difference with the standard bifurcation theory in which the parameters vary continuously (consider, for example, the period doubling bifurcation in the logistic map).

This paper complements the results in the interesting works of Klic and Pokorny [13], [14]. We show examples of "chaos+ chaos=order" in the periodic combination of discrete dynamical systems and in [13] and [14] it is shown examples of two vector fields $v$ and $w$ with chaotic and strange attractors such that when periodically combined they give rise to a single point attractor. $v$ and $w$ are not free but they are topologically conjugate via an involutory diffeomorphism $G$ (by the term involutory we mean that $G = G^{-1}$), on the contrary in our work the discrete dynamics $A_1$ and $A_2$ are not conjugate. In this sense our results are more general than those of Klic and Pokorny.

In ending the introduction an important comment concerning the relevance of our results is in order. The study of Parrondian phenomena is a very recent area of research and therefore occurrence of these phenomena in the real world is still object of discussion. Several applications to physics, biology and economy have been proposed [15-19]. Furthermore the alternation of continuous dynamical systems (vector fields) arises at the modelling of tubular catalytic reactors, kicked rotators or blinking vortex flows. Note that in Nature there are many different interactions and therefore systems do not evolve according to a unique dynamics. It is reasonable that the evolution of certain physical, biological or economical complex systems can only be explained by the combination of different dynamics. It is at this step where Parrondian phenomena could play a role and yield behaviours that each separated dynamics does not allow.

## 2. Periodic combination of chaotic quadratic maps

Let us consider two dynamics $A_1$ and $A_2$ defined by the one-dimensional real Mandelbrot maps

$$A_1: \quad x_{n+1} = x_n^2 + c_1,$$
$$A_2: \quad x_{n+1} = x_n^2 + c_2.$$

It is well known [20, 21] that when $-2 \leq c \leq 0.25$ the invariant set under iteration of the real Mandelbrot map, that is, the set of initial conditions yielding bounded orbits, is the interval $\left[ -\frac{1+\sqrt{1-4c}}{2}, \frac{1+\sqrt{1-4c}}{2} \right] \subset [-2,2]$. When $c < -2$ [20, 21] the invariant set is a zero Lebesgue measure Cantor set in $[-2,2]$. When $c > 0.25$ all the orbits escape to infinity.

We define the alternated dynamics $(A_1 A_2)$ as

$$\left( A_1 A_2 \right): \quad \begin{cases} x_{n+1} = x_n^2 + c_1 \text{ when } n \text{ is even} \\ x_{n+1} = x_n^2 + c_2 \text{ when } n \text{ is odd}, \end{cases} \tag{4}$$

and in order to study the alternated dynamics $(A_1 A_2)$ we also define the auxiliary dynamics $B$

$$B: \quad x_{n+1} = f(x_n) = \left( x_n^2 + c_1 \right)^2 + c_2.$$

Note that there exists a simple 1-1 correspondence among the orbits of $(A_1A_2)$ and $B$ because the period of an orbit of $(A_1A_2)$ is just twice the period of the corresponding orbit of $B$.

If the parameters $c_1$ and $c_2$ in Eq. (4) satisfy the constraint $c_1^2 + c_2 = 0$ it is easy to see that the point $x = 0$ is a superstable fixed point of $B$ and also it is one of the points of a period-2 orbit of $(A_1A_2)$. We say that this orbit is superstable because it contains the point $x = 0$.

Let us now show that a similar conclusion holds when $c_1^2 + c_2 = e$, $e$ standing for a small enough real number. Since $f(x) = x^4 + 2c_1x^2 + e$, it follows that the equation defining the fixed points of $B$ is $g(x, e) = x^4 + 2c_1x^2 - x + e = 0$. As $\left(\dfrac{\partial g}{\partial x}\right)_{(x=0, e=0)} = -1$ and $g$ is an analytic function we can apply the implicit function theorem in order to obtain the value of the fixed point of $B$ close to 0 in terms of powers of $e$. Indeed the Taylor expansion of the analytic function $x(e)$ is convergent when $e$ is small and is given by $x(0) + x'(0)e + O(2)$. When $e = 0$ then the stable fixed point of $B$ is $x(0) = 0$. On account of the implicit function theorem, we have that $x'(0) = \left(-\dfrac{\partial g/\partial e}{\partial g/\partial x}\right)_{(0,0)} = 1$ and therefore we get that $x(e) = e + O(2)$. Now $\left|f'(x)\right|_{x=e+O(2)} = \left|4c_1e + O(2)\right| \ll 1$ whenever $c_1$ is in the interval $[-2, 0.25]$ and $e$ is small enough. Summarizing, when the condition

$$c_1^2 + c_2 = e \qquad (5)$$

is fulfilled then the dynamics $(A_1A_2)$ possesses a 2-periodic stable orbit that has a point near $x = 0$. The smaller $e$ is the more stable the orbit is and in the limit, $e = 0$, it is superstable. When $c_1 \approx -1.4$ (see Example 1 below) then the upper bound of $|e|$ is 0.09. Indeed, since the fixed point of the dynamics $B$ must be stable, we have the following condition $\left|f'(x)\right| = \left|4x^3 - 5.6x\right| < 1 \Rightarrow |x| < 0.18$. Note now that $x \approx -0.18$ is a fixed point of $f(x)$ if $e \approx -0.09$ and when $x \approx 0.18$ the corresponding value is $e \approx 0.27$.

Therefore if $e \in (-0.09, 0.27)$ we get fixed points around the origin which are stable, in particular when $|e| < 0.09$, as we desired to prove.

The presence of a stable periodic orbit guarantees that the dynamics $(A_1 A_2)$ is ordered. Indeed no definition of chaos admits the existence of stable periodic orbits. Furthermore note that dynamics $B$ satisfies the following properties:

(1) There exists an interval $I = [-\boldsymbol{I}, \boldsymbol{I}]$, $\lambda$ a small enough positive real number, which is invariant under $f$.

(2) $f$ has a single local maximum at $x = 0$ ( $f''(0) = 4c_1 < 0$ ).

(3) The Schwarzian derivative of $f$ is negative for all $x \in I - \{0\}$.

Then by Guckenheimer's theorem [22] the fact that $B$ possesses a stable periodic orbit implies no sensitivity to initial conditions, a fundamental ingredient of any definition of chaos.

Note that nothing has been said about the single dynamics $A_1$ and $A_2$. The question is therefore: can we find parameters $c_1$ and $c_2$ satisfying Eq. (5) such that dynamics $A_1$ and $A_2$ are chaotic? It is well known that when $c \in [-2, 0.25]$ the chaotic region of the real Mandelbrot map is the interval $[-2, c_{MF}]$, where $c_{MF} = -1.401155189...$ is the parameter value of the Myrberg-Feigenbaum point. Since $c_1$ and $c_2$ must satisfy Eq. (5) it is immediate that they must belong to the intervals (when $e = 0$ )

$$-\sqrt{2} \leq c_1 \leq c_{MF},$$
$$-2 \leq c_2 \leq -c_{MF}^2.$$

The only points in the chaotic region which are chaotic in the strict mathematical sense are the Misiurewicz points [23]. They verify the following properties:

(1) Existence of infinite periodic orbits of period $2^n$ ( $n \geq 0$ ). All of them are unstable (there not exist stable periodic orbits) [24].

(2) Sensitive dependence to initial conditions [24] which is manifested in the positivity of the Lyapounov exponent [25, 26].

(3) Ergodicity in certain subset of the invariant set [24].

Note that when $c \leq -2$ the real Mandelbrot map is chaotic in the sense of Devaney [20, 21] (to our knowledge the strongest definition of chaos).

We conclude that if $c_1$ and $c_2$ are Misiurewicz points satisfying condition (5) then the two chaotic dynamics $A_1$ and $A_2$ yield an ordered dynamics $(A_1 A_2)$. In the following two examples we show that these points indeed exist and therefore that "chaos + chaos = order" is possible.

## 3. Example 1

The most representative chaotic point in the interval $-\sqrt{2} \leq c_1 \leq c_{MF}$ and near the parameter value $-\sqrt{2}$ corresponds to $c_1 = -1.407405118...$ It is the Misiurewicz point $M_{9,4}$ separating the chaotic bands $B_2$ and $B_3$ [23]. The corresponding orbit, obtained iterating the critical point $x = 0$, is unstable and has preperiod 9 and period 4. As is well known the critical polynomials of the real Mandelbrot map $x_{n+1} = x_n^2 + c$ are $P_0 = 0$, $P_1 = c$, $P_2 = c^2 + c$, $P_3 = (c^2 + c)^2 + c$ ... and we can obtain the parameter value of $M_{9,4} = c_1$ by solving $P_{9+4} = P_9$. This equation has many solutions, but $c_1$ is the zero of $P_{9+4} - P_9$ located near $-\sqrt{2}$ and we can easily determine this parameter value with the necessary precision. It is enough to use a 80 bits floating point precision (about 16 decimal digits) in the simulations. The reader can see in Fig. 3 the ergodicity regions of the invariant interval $[-1.787402470...,1.787402470...]$ and in Fig. 4 the graphical iteration of the initial condition $x = 0$. The Lyapounov exponent can be estimated via Shaw formula [25] obtaining the positive value $\Lambda = 0.086$. It is clear that $A_1$ is a chaotic dynamics for all practical purposes and also in a strict mathematical sense.

The parameter value $c_2 = -2$ is the Misiurewicz point $M_{2,1}$ in the chaotic band $B_0$ [23]. In this case the ergodicity region is the whole invariant interval $[-2,2]$ (see the histogram of the orbit in Fig. 5 and the graphical iteration of the initial point $x_0 = 0$ in Fig. 6). Note that these figures have been obtained using the parameter value $c_2 = -1.999999999...$ instead of $c_2 = -2$ to simulate the necessary instability. The Lyapounov exponent is the positive value $\Lambda = \ln 2$. Therefore $A_2$ is a chaotic dynamics. In fact it is more chaotic than $A_1$ because it is chaotic in the sense of Devaney.

The parameter $\boldsymbol{e} = c_1^2 + c_2 = -0.019210834...$ is small enough and therefore, according to the discussion above, the dynamics $B$ must be ordered. The only stable fixed point of $B$ is $-0.020379747...$, near the origin, and the dynamics $(A_1 A_2)$ has the 2-periodic stable orbit $\{-1.406989784..., -0.020379747...\}$. The corresponding Lyapounov exponent is negative $\Lambda = -2.1$. Note that the invariant set $\Omega$ under iteration of $(A_1 A_2)$ is not an interval but it contains at least the interval around the origin $[-0.3, 0.3]$. In Fig. 7 we show the histogram of the combined dynamics and in Fig. 8 the graphical iteration of the initial point $x_0 = -0.3$. There is no doubt that the alternated dynamics is ordered.

Note that in the interval $[-1.407405118..., -1.401155189...]$ there are infinite Misiurewicz points separating the chaotic bands $B_i$ and $B_{i+1}$ ($i \geq 3$). Taking $c_2 = -2$ all these Misiurewicz points $c$ satisfy the condition $|c^2 + c_2| = |\boldsymbol{e}| < 0.037$ and therefore, on account of the results in section 2, there exist infinite examples of "chaos + chaos = order".

## 4. Example 2

This example possesses more mathematical than physical interest. Take $c_1 = -2$ and $c_2 = -4$. The invariant set of $A_1$ is the interval $[-2, 2]$ but the invariant set of $A_2$ is a zero Lebesgue measure Cantor set $\Omega_2$. Both dynamics are Devaney chaotic in their respective domains. The Lyapounov exponent of $A_1$ is positive and the Lyapounov exponent of $A_2$ tends to $+\infty$ when growing the number of iterations. On the contrary the dynamics $(A_1 A_2)$ possesses Lyapounov exponent tending to $-\infty$ with the number of iterations, and since $c_1^2 + c_2 = 0$ the orbit $\{0, -2\}$ is superstable and $(A_1 A_2)$ is an ordered dynamics.

The main differences between this example and Example 1 are that the single dynamics $A_1$ and $A_2$ are both chaotic in the sense of Devaney and that the set $\Omega_2$ is not an interval and therefore almost all orbits of $A_2$ escape to infinity.

## 5. Final remarks

In this paper we have shown that the periodic combination of two chaotic dynamics of the form $x^2 + c$ for certain values of the parameter $c$ results in an ordered dynamics. This "chaos + chaos = order" phenomenon is analogous to the original Parrondo's paradox "losing + losing = winning" and is obtained by exploiting Parrondo's philosophy of alternation. Different paradoxical phenomena can be studied by alternating non linear deterministic dynamics of dimension 1 or 2.

Recall that the real Mandelbrot map $x_{n+1} = x_n^2 + c$ is topologically conjugate to the logistic map $x_{n+1} = m x_n (1 - x_n)$ and therefore the phenomenon "chaos + chaos = order" also arises in the combination of logistic maps. For instance the values of $m_1$ and $m_2$ associated to the values of $c_1$ and $c_2$ in Example 1 are $m_1 = 3.574804938...$ and $m_2 = 4$. The logistic maps $\tilde{A}_i : x_{n+1} = m_i x_n (1 - x_n)$ $(i = 1, 2)$ are chaotic but the alternated map $(\tilde{A}_1 \tilde{A}_2)$ is ordered. Note however that $(\tilde{A}_1 \tilde{A}_2)$ is not topologically conjugate to $(A_1 A_2)$ since the first one has a 8-periodic stable orbit (see Fig. 9) and the second one has a 2-periodic stable orbit.

The following remark is interesting. One can interpret the model presented as a periodic switching of a single parameter $c$. Consider an arbitrary period $T$ for this switching. For $T$ large, the system has time enough to reach the stationary regime of each dynamics which is a chaotic attractor. However, decreasing $T$ this attractor becomes a periodic orbit. We would have a transition from order to chaos varying the dynamical parameter $T$. Let us illustrate this phenomenon with an example. Consider the logistic maps $\tilde{A}_1 : x_{n+1} = m_1 x_n (1 - x_n)$ and $\tilde{A}_2 : x_{n+1} = m_2 x_n (1 - x_n)$ with $m_1 = 3.574804938...$ and $m_2 = 4$. As it is shown in the preceding paragraph these dynamics are chaotic but their switching with $T = 1$: $(\tilde{A}_1 \tilde{A}_2)$ is ordered (it possesses a 8-periodic stable orbit). If now we consider the switchings with periods $T = 2$: $(\tilde{A}_1 \tilde{A}_1 \tilde{A}_2 \tilde{A}_2)$ and $T = 3$: $(\tilde{A}_1 \tilde{A}_1 \tilde{A}_1 \tilde{A}_2 \tilde{A}_2 \tilde{A}_2)$ we see that the transition from order to chaos is abrupt (see Fig. 9) and even when $T = 2$ the switched dynamics is chaotic.

It goes without saying that chaotic dynamics $A_1$ and $A_2$ studied in section 2 can be combined in other deterministic manners (for example $A_1A_1A_2A_1A_1A_2...$) or even randomly (regarding the study of Lyapounov exponents for stochastic combination of dynamics it is worth mentioning the interesting work of Kocarev [27]). Whether paradox occurs or not in all these cases remains an open problem. Note that condition (5) will change and therefore Misiurewicz points will have to satisfy other conditions. In the random alternation we do not know how to accomplish a theoretical study.

**Acknowledgments**

D.P.-S. is supported by an FPU grant from Ministerio de Educación, Cultura y Deportes (Spain). M.R. is indebted to the Ministerio de Ciencia y Tecnología for financial support through 2001-0586. The authors also thank the referees for his/her useful remarks.

*Corresponding author. Email address: dperalta@fis.ucm.es

**Note added in proof**

The authors have only recently been aware of the existence of an interesting paper which studies the topological entropy of the compositions of commuting dynamics, J.S. Canovas and A. Linero, Nonlinear Anal. 51 (2002) 1159. In this work it is proved that the topological entropy of the combination is lower or equal than the sum of the single entropies for the case in which the dynamics commute. Note that the real Mandelbrot maps that we consider do not commute among them and therefore this general result does not apply.

**Figure captions**

**Fig. 1.** Representation of the surfaces $f = 1$ and $\Lambda = 1$ in the parameters space $(p_1,\ p_2,\ p)$ with $s = 0.4$. Note that $p < \dfrac{1}{2}$ in all the diagram and both surfaces intersect each other, delimiting a region within Parrondo´s paradox occurs, as far as conditions (1) and (2) hold there.

**Fig. 2.** Expected gain curves obtained playing separately with games $A_1$ and $A_2$, combining them in the way $\{a,b\} = \left(A_1^a A_2^b\right)^{100/(a+b)}$ (with values of the parameters $p = 0.495$, $p_1 = 0.095$, $p_2 = 0.745$) and combining them randomly with $s = 0.4$. Notation $\left(A_1^a A_2^b\right)^{100/(a+b)}$ means that we first iterate game $A_1$ $a$ times, then game $A_2$ $b$ times and repeat the same alternation sequence $100/(a+b)$ times. All the curves have been obtained averaging over 50,000 trials.

**Fig. 3.** Histogram of the orbit of the real Mandelbrot map for the parameter value $c_1 = -1.407405118....$

**Fig. 4.** Graphical iteration in the real Mandelbrot map for the parameter value $c_1 = -1.407405118....$

**Fig. 5.** Histogram of the orbit of the real Mandelbrot map for the parameter value $c_2 = -2$.

**Fig. 6.** Graphical iteration in the real Mandelbrot map for the parameter value $c_2 = -2$.

**Fig. 7.** Histogram of the orbit in the alternated dynamics $\left(A_1 A_2\right)$ $(A_1: x_{n+1} = x_n^2 - 1.407405118..., A_2: x_{n+1} = x_n^2 - 2)$.

**Fig. 8.** Graphical iteration in the alternated dynamics $\left(A_1 A_2\right)$ $(A_1: x_{n+1} = x_n^2 - 1.407405118..., A_2: x_{n+1} = x_n^2 - 2)$. The initial point is $x_0 = 0.3$.

**Fig. 9.** Logistic maps $\tilde{A}_1: x_{n+1} = m_1 x_n (1 - x_n)$ and $\tilde{A}_2: x_{n+1} = m_2 x_n (1 - x_n)$ with $m_1 = 3.574804938...$ and $m_2 = 4$. a) Histogram with switching period $T = 1$: $(\tilde{A}_1\tilde{A}_2)$. b) Histogram with switching period $T = 2$: $(\tilde{A}_1\tilde{A}_1\tilde{A}_2\tilde{A}_2)$. c) Histogram with switching period $T = 3$: $(\tilde{A}_1\tilde{A}_1\tilde{A}_1\tilde{A}_2\tilde{A}_2\tilde{A}_2)$.

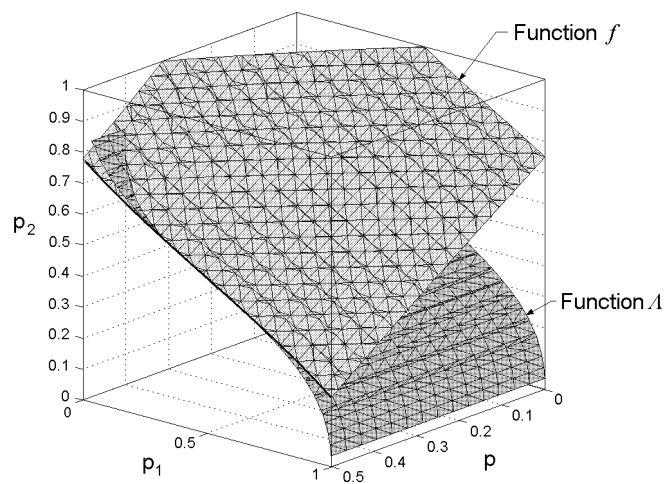

Fig. 1

J. Almeida, D. Peralta-Salas, and M. Romera

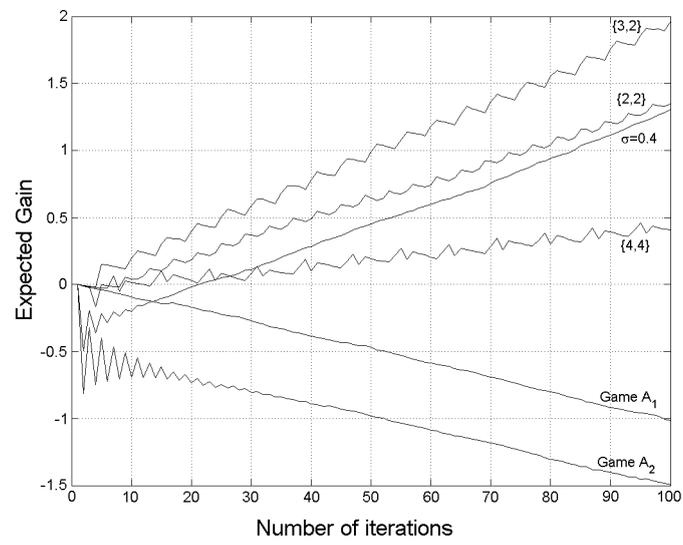

Fig. 2

J. Almeida, D. Peralta-Salas, and M. Romera

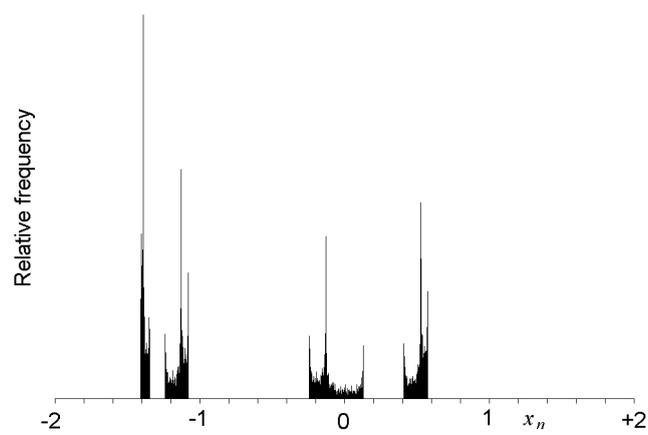

Fig. 3

J. Almeida, D. Peralta-Salas, and M. Romera

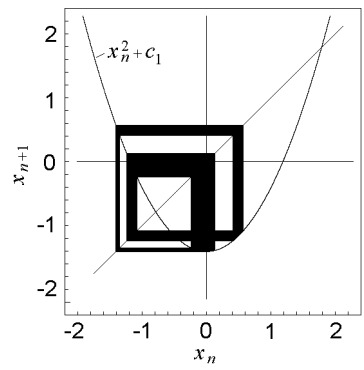

Fig. 4

J. Almeida, D. Peralta-Salas, and M. Romera

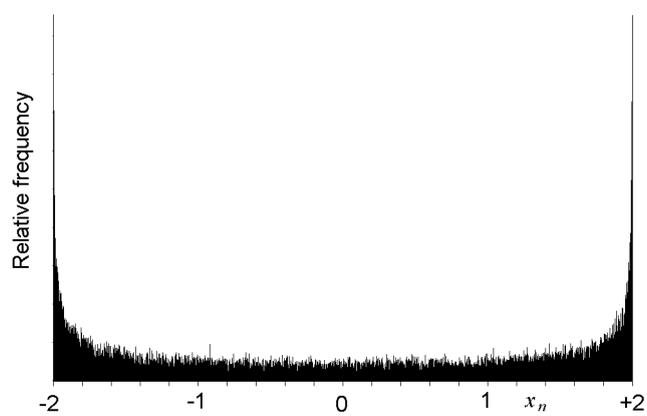

Fig. 5

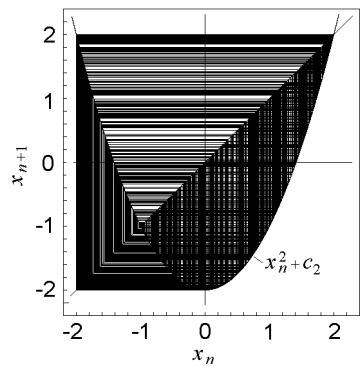

Fig. 6

J. Almeida, D. Peralta-Salas, and M. Romera

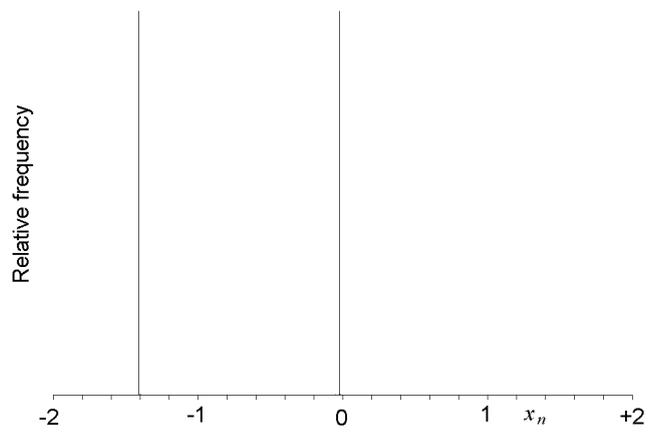

Fig. 7

J. Almeida, D. Peralta-Salas, and M. Romera

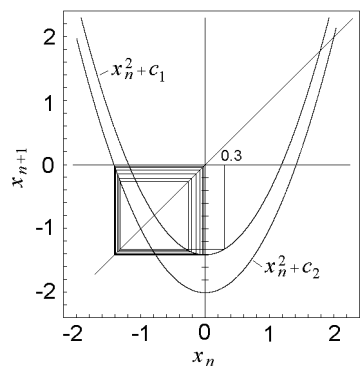

Fig.8

J. Almeida, D. Peralta-Salas, and M. Romera

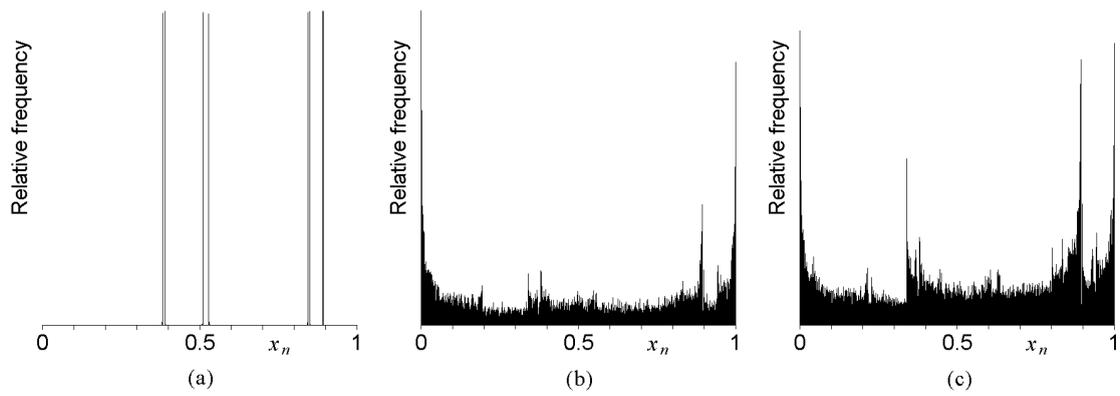

Fig.9